\documentclass[twocolumn]{article}

\usepackage[applemac]{inputenc}
\usepackage[T1]{fontenc} 
\usepackage[ngerman,english]{babel}
\usepackage{amsmath,amssymb,amsthm}
\usepackage{natbib}
\usepackage{setspace}

\theoremstyle{definition}
\newtheorem{definition}{Definition}

\newtheorem{proposition}[definition]{Proposition}
\newtheorem{definition/proposition}[definition]{Definition/Proposition}

\begin{document}

\title{Gerrymandering Individual Fairness}

\author{Tim Räz\footnote{Institute of Philosophy, University of Bern, Switzerland. tim.raez@posteo.de}}


\maketitle

\begin{abstract}
Individual fairness, proposed by Dwork et al., is a fairness measure that is supposed to prevent the unfair treatment of individuals on the subgroup level, and to overcome the problem that group fairness measures are susceptible to manipulation, or gerrymandering. The goal of the present paper is to explore the extent to which it is possible to gerrymander individual fairness itself. It will be proved that gerrymandering individual fairness in the context of predicting scores is possible. It will also be argued that individual fairness provides a very weak notion of fairness for some choices of feature space and metric. Finally, it will be discussed how the general idea of individual fairness may be preserved by formulating a notion of fairness that allows us to overcome some of the problems with individual fairness identified here and elsewhere.
\end{abstract}

\section{Introduction}

The fair-ML debate distinguishes different kinds of fairness, among them measures of group fairness and individual fairness \citep{kearn2018, mitch2021}. Group fairness measures are formulated on an aggregate level. It has been argued \citep{dwork2012} that group fairness measures allow for gerrymandering, that is, it is possible to manipulate predictors such that they satisfy measures of group fairness, yet the predictions contradict intuitions of fairness. As a remedy, \citet{dwork2012} proposed individual fairness, a measure that prevents the unfair treatment of individuals within groups to a certain extent. Informally, individual fairness (IF) requires that similar people should be treated similarly. This idea is captured by a formal condition on predictors, formulated in terms of metrics between individuals and predictions.

The goal of the present paper is to explore the extent to which it is possible to gerrymander IF itself. I will prove that it is possible to gerrymander IF, in particular settings, both with respect to groups and with respect to individuals. I will also argue that IF provides a very weak notion of fairness. One of the main problems of IF is already apparent in the informal statement above: A lot hinges on what we mean by ``similar''. How do we measure the similarity between people? What features of individuals are relevant to determine similarity, and what is an appropriate way of measuring it? \citet{dwork2012} propose that this is not a problem, but a feature of IF: choosing an appropriate metric is part and parcel of spelling out what we mean by fairness in a particular context. 

IF is a formal notion of fairness, as opposed to a substantive one. Few properties of IF can be explored on the basis of the definition of IF alone. IF only becomes a substantive notion once we consider it in a particular setting, i.e., by choosing a particular feature space for individuals, metrics, and a predictor. The same is true if one wants to show that gerrymandering IF is possible. This is why I will explore gerrymandering IF in the context of predicting scores, that is, real-valued predictions (Sec. \ref{sec:ger_pred}). This is a specific context that will bring some problems of IF to the fore. I will also explore gerrymandering features and metrics (Sec. \ref{sec_metric_features}). Here the goal is to explore the space of possibilities spanned by IF. Finally, I will discuss how we might preserve the general idea of individual fairness by formulating a notion of fairness that allows us to overcome some of the problems with IF identified here (Sec. \ref{sec:leibnitz}).

\section{Background and Related Work}
\label{sec:background}

The discussion of different kinds of group fairness picked up speed in reaction to the publication of \citet{angwi2016}, which examined COMPAS, a risk assessment instrument, and found that COMPAS violates one kind of group fairness. There are now many surveys of fair-ML \citep{mitch2021}; \citet{baroc2019} provide a regularly updated, book-length treatment of fair machine learning (with a focus on measures of group fairness); \citet{kearn2019} provide a book-length introduction to ethical ML algorithms. The problem of gerrymandering statistical parity (also known as demographic parity, independence), a notion of group fairness, was noted in \citet{dwork2012}. In fact, the possibility of gerrymandering statistical parity was one of the main motivations of Dwork et al. to propose individual fairness. It has since been pointed out \cite{raez2021} that it is also possible to gerrymander other measures of group fairness, such as sufficiency (calibration) and separation (equalized odds).

Individual fairness was first proposed in \citet{dwork2012}. For the formal definition of individual fairness, we need a random variable $X$ encoding the set of ``individuals'', and a random variable $\hat{Y}$ encoding the prediction space. The predictor $M$ is randomized, meaning that it maps to distributions over $\hat{Y}$.

\begin{definition}
\textbf{(IF):} Let $D, d$ be metrics. A randomized mapping $M:X \rightarrow \Delta(\hat{Y})$ satisfies \textbf{individual fairness} if it satisfies the $(D, d)$-Lipschitz property, i.e., if for $p, q \in X$, we have: $D(Mp, Mq) \leq d(p,q)$.
\end{definition}

Note that, strictly speaking, $D$ and $d$ are not metrics, but only pseudo-metrics, because metrics presuppose that $d(p, q) = 0$ iff. $p = q$, while Dwork et al. (p. 1) allow for the case that $d(p,q) = 0$ for $p \neq q$. I will work with pseudo-metrics throughout the paper.

Other variants of IF have been proposed. \citet{fried2016} require that if individuals are $\epsilon$-close (in so-called Construct Space), then they should be mapped to $\epsilon$'-close predictions (in so-called Decision Space). \citet{shari2019} define a similar notion of \emph{average individual fairness}. \citet{kearn2018} investigate the possibility of preventing gerrymandering group fairness measures by considering measures on rich subgroups, thus interpolating between group and individual fairness. 

It is noted in \citet{dwork2012} that IF is a strong requirement which may be hard to enforce. There have been several works addressing the problem of constructing IF predictors, e.g., through elicitation of the metric \citep{ilven2019,mukhe2020}. A framework for enforcing a fairness measure related to IF, which preserves order structure, is proposed in \citet{jung2019}. There are some proposal to enforce IF in-training \citep{yuroc2019,yuroc2020,vargo2021}, and in post-processing \citep{peter2021}.

There have been few critical examinations of IF from a conceptual or philosophical point of view; \citet{binns2018} provides for a useful overview of philosophical work relevant to fair-ML. \citet{fleish2021} provides a detailed critical examination of IF. Fleisher describes two ways of gerrymandering IF: Universal rejection (constant treatment), and increasing all scores by a fixed amount. Fleisher does not explore further possibilities of gerrymandering IF predictors, and does not examine general metric spaces, a gap filled by the present paper (Sec. \ref{sec:ger_pred}). Note also that Fleisher does not specifically discuss the question whether the cases of gerrymandering discussed by him constitute discrimination. According to a working definition \citep{altma2020}, discrimination means that people from a socially salient group are put at a relative disadvantage due to their group membership. Fleisher discusses issues that are beyond the scope of the present paper, such as determining a fairness metric through elicitation of moral judgements, and the problem that a fairness metric presupposes commensurability of predictions, which may be violated if values are incommensurable. \citet{binns2020} has examined the relation between group fairness and IF, arguing that the conflict between the two kinds is only apparent. One argument for this thesis proposed by Binns is that IF is a notion based on statistical generalization, as opposed to a truly individualized notion of fairness, and should thus be regarded as kind of group fairness, lumping together individuals with the same features. This point will be discussed below (Sec. \ref{sec_metric_features}).

\section{Gerrymandering Predictors}
\label{sec:ger_pred}

\subsection{Idea and Strategy}

The goal of this section is to investigate the possibility of gerrymandering predictors that satisfy IF. To make the general definition of IF workable, I will use a particular choice of the metric $d$ on the prediction space $\hat{Y}$. I work with point predictions $M:X \rightarrow \hat{Y}$ that take the form of a score, that is, predictions $\hat{y} \in \hat{Y}$ are in a linearly ordered set, such as $\mathbb{R}$. I will use the Euclidean distance $d(p, q) = |p-q|$ on $\mathbb{R}$ as the metric between predictions; I will not make use of the metric on the space of individuals $X$. To give an example for this setting, the prediction could be the assignment of a score $\hat{y} \in \hat{Y}$ to individuals in $X$, where the score captures how well they are suited for a job.

To make the following ideas more easily graspable, it is useful to think of the predictions in $\hat{Y}$ as capturing a ground truth in $Y$ with the same structure as $\hat{Y}$. Thus, the setting is very similar to standard supervised learning, where we try to approximate a ground truth $Y$ with a predictor $\hat{Y}$. With this choice, one can also think of the metric between individuals as being at least partially determined by their ``true ability'' $Y$.

Informally, the idea behind this section is as follows. I assume that we are given a predictor $M$ that assigns scores to individuals, e.g., their suitability for a job. This means that it orders these individuals in some way. I assume further that this predictor satisfies IF. I also assume that the scores assigned to individuals are fair in a substantive sense, in that the predictor does not discriminate against different groups or individuals because of their group membership. I will then show that it is possible to manipulate the predictor such that IF is still provably satisfied, but that specific groups or individuals are assigned a different score that puts them at a disadvantage, which constitutes discrimination. This shows that IF allows for fairness gerrymandering.

Formally, I will assume that we have determined that a certain predictor $M:X \rightarrow \hat{Y}$ satisfies IF. Now, if we only aim to preserve IF, we also have to accept other predictors $M'$ that can be constructed from $M$  and that also comply with IF, but violate fairness in a substantive sense. The main tool for constructing gerrymandered predictors $M'$ from fair predictors $M$ is to transform $M$ into $M'$ such that the Lipschitz property is preserved. This can be done by finding a mapping $\phi:\hat{Y} \rightarrow \hat{Y}$ that is Lipschitz. Assume we have a IF predictor $M:X \rightarrow \hat{Y}$. If we compose a Lipschitz mapping $\phi$ with the predictor $M$, this will yield a predictor $M':X \rightarrow \hat{Y}$ with $M' := \phi \circ M$, which also satisfies IF, but may be discriminating.

\subsection{General Non-expansive Maps}
\label{sec:non-expansive}

To start, it is useful to examine what kinds of maps $\phi$ on metric spaces are Lipschitz. First, note that a map $\phi$ will preserve the Lipschitz property if and only if it is non-expansive, that is, if it does not increase the distance between points $p, q \in \hat{Y}$:

\begin{definition}
Let $(Y, d)$ be a metric space. A mapping $\phi:Y \rightarrow Y$ is \textbf{non-expansive} if for all $p, q \in Y$ it holds $d\big(\phi (p),\phi (q)\big)\ngeq d\big(p,q\big)$.
\end{definition}

It is useful to distinguish two kinds of mappings that are non-expansive, namely isometries and contractions. First, let us examine isometries. Isometries preserve the distance between individuals:

\begin{definition}
Let $(Y, d)$ be a metric space. A mapping $\phi:Y \rightarrow Y$ is an \textbf{isometry} on $Y$ if for all $p, q \in Y$ it holds $d(p, q)=d \big(\phi(p), \phi(q)\big)$.
\end{definition}

There are two kinds of isometries that preserve Euclidean distances on the line; cf. \citet[Sec.1.5]{petru2021}: reflections and translations. 

A translation shifts the scores of a fair predictor $M:X \rightarrow \hat{Y}$ by a fixed amount $c \in \mathbb{R}$ through a mapping $\phi:\hat{Y} \rightarrow \hat{Y}$, with $\phi(\hat{y}) = \hat{y} +c$. It has been pointed out in the literature \citep[Sec.3.1.]{fleish2021} that translations preserve the Lipschitz property. Note, though, that this only works in the particular setting chosen here; translations need not be isometries in non-Euclidean metric spaces. A translation need not be discriminatory, because it applies equally to all individuals, and does not necessarily target any group or individual. For discrimination, it is necessary that people from a socially salient group are put at a relative disadvantage due to their group membership \citep{altma2020}. However, translations can be used in a targeted manner to discriminate under certain conditions. Assume that some group $A$ has a high concentration (is overrepresented) in an interval $[t, t']$ above a threshold $t$, such that $\hat{y} \geq t$ is the positive class (e.g., suitable for the job), and $\hat{y} <t$ is the negative class (not suitable). If a gerrymanderer now wishes to hurt group $A$, they can apply a translation $\phi$ with $\hat{y} \mapsto \hat{y} - t'$, pushing people with scores in $[t, t']$ below the threshold. The new predictor still satisfies IF with respect to the score, but disproportionately hurts people from group $A$, concentrated in the interval. Similarly, the gerrymanderer can lift a group $B$ concentrated below a threshold up.

The possibility of gerrymandering in the above examples depends on two things: A particular distribution of different groups over scores, and a particular utility of putting a certain group at a disadvantage. Gerrymandering as just described only makes sense if the gerrymanderer's utility of, say, pushing people from group $A$ below a threshold needs to be higher than the utility of keeping people from other groups, which may also have scores in the interval $[t, t']$.

Turning to reflections, assume that we have a predictor $M:X \rightarrow \hat{Y}$, and a threshold $t \in \hat{Y}$, such that $\hat{y} \geq t$ is the positive class, and $\hat{y} <t$ is the negative class. A reflection with respect to the score, $\phi:\hat{Y} \rightarrow \hat{Y}$, $\hat{y} \mapsto -\hat{y} + 2t$ will preserve IF with respect to scores: If $M$ is an IF predictor, then so is $M' := \phi \circ M$. A reflection with respect to the score will map all individuals that are in the positive class according $M$ to the negative class, and vice versa. Of course, reflections could also be used in a more targeted manner to flip the scores using any point of reflection. As mentioned before, a reflection need not constitute discrimination, because it does not necessarily target a specific group, and may not align with a gerrymanderer's utility.

Let us now turn to contractions. A contraction is a map that lowers the distance between any two points:

\begin{definition}
Let $(Y, d)$ be a metric space. A mapping $\phi:Y \rightarrow Y$ is a \textbf{contraction} if there is a constant $k \in [0, 1)$ such that for all $p, q \in Y$ it holds $d\big(\phi(p),\phi(q)\big) \leq k\cdot d(p,q)$.
\end{definition}

A first example of a contraction is a map $\phi$ that sends all individuals to a single value $\hat{y}^* \in \hat{Y}$. This is a contraction because it reduces the distance between all predictions to $0$. The fact that constant predictors are Lipschitz was already noted in \citet{dwork2012} and also in \citet{fleish2021} in the form of \emph{Universal Rejection}: If a score measures suitability for college, it is IF to assign the same, low score to all applicants, and thus reject them all. Fleisher notes that this may be considered unfair to those who would be suitable for college, but are rejected. Note that universal rejection need not constitute discrimination in the sense that people are treated differently because of their group membership. Note also that constant predictions may be unfair for other reasons. For example, if we want to assign income tax rates to individuals, and use a constant predictor, i.e. a flat income tax, this will effectively yield a degressive tax scheme, because a fixed percentage of income does not have the same utility for people with high incomes as for people with low incomes.

\subsection{Local Contractions}

Now I turn to a more targeted kind of gerrymandering based on \emph{local} contractions. Assume that a certain group $A$ is has a high concentration in an interval $[t, t']$ of scores. The predictions in this interval can be contracted to a point $t^*$ in that interval, shifting scores below $t$ up with $\hat{y} \mapsto \hat{y} + (t^* - t)$, and scores above $t'$ down with $\hat{y} \mapsto \hat{y} - (t' - t^* )$. Of course, such a locally constant map can be applied to several intervals at the same time, and it can also be used in combination with thresholds to yield similar effects as in the case of translations, discussed above. Local contractions can be considered to be discriminatory against the targeted group $A$ because the difference between of members of this group, mirrored by the different scores, are eliminated (at least in the interval in question), such that members of this group are made to look more similar in contrast to other groups.

In general, contractions can be chosen such that any interval of the score is assigned its own rate of contraction, where intervals can be chosen based on the concentration of group membership in the respective intervals, so as to maximize utility for the gerrymanderer. As in the case of translations, local contractions will also affect the individuals above and below the contracted interval, because the map needs to be non-expansive for all pairs of individuals. Whether a local contraction is in the interest of the gerrymanderer is a function of their utility: The change of scores of the individuals that are ``pulled down'' or ``pushed up'' should lead to a smaller loss of utility than the gain from the local contraction.

\subsection{Local Reflections (Folding Attack)}

In this section, I present an attack that is a sort of local reflection. The manipulated predictor yields a kind of gerrymandering that is very similar to an attack against statistical parity proposed by \citet{dwork2012}, which was supposed to \emph{support} IF as a notion of fairness.

In order to gerrymander the predictor, define the following map:

\begin{definition}
A \textbf{folding} map $\phi:\mathbb{R} \rightarrow \mathbb{R}$ for an interval $[a, b] \subset \mathbb{R}$, $a < b$ is defined as follows:

\begin{equation*}
\phi(x) = \begin{cases}
x - 2(b-a) & \text{if } x > b, \\
-x + 2a & \text{if } x \in [a, b] \\
x & \text{if } x < a.
\end{cases}
\end{equation*}

\end{definition}

A folding map adds a fold at the points $a, b$, such that scores between $a$ and $b$ are reversed, while the scores above $b$ are translated down by twice the distance between $a$ and $b$, and the scores below $a$ are left unchanged. Folding maps can be used for gerrymandering because they preserve IF, a proof is provided in appendix \ref{sec:folding_proof}:

\begin{proposition}
Folding maps are non-expansive with respect to the Euclidean metric.
\label{prop:folding}
\end{proposition}

How can a folding map be used to gerrymander? Essentially, a folding map is a local reflection, and makes it possible to put people with a lower score above people with a higher score, without reversing the order of all scores, as in a reflection. This means that you can target a certain interval of the entire range of the score. By applying a folding map to an interval $[a, b]$ in which group $A$ is highly concentrated, you can adversely affect that group by putting people \emph{predominantly from that group} with low scores before people \emph{predominantly from that group} with high scores. This yields a version of what Dwork et al. call \emph{Self-fulfilling prophecy}: By putting worse candidates of one group in front of more suitable ones, people from that group end up with scores that are not reflective of their abilities, which can later be used as a justification for future discrimination. As with contractions, it is possible to apply folding maps in sequence to different intervals. It is also possible to apply the folding map to an interval directly above a threshold, if the group $A$ is concentrated in that interval, to obtain unfavorable prediction for that group.

Applying a folding map comes at the price of applying a contraction to the regions separated by the fold, that is, individuals above and below the folding interval will receive scores that are closer together. Using a folding attack to an interval relies both on a certain distribution of groups over scores and on the utility of doing / preventing harm to different groups. For example, a folding attack may be in the interest of an attacker even if the folding interval contains two groups in equal distribution, if the utility of hurting one group is bigger than preventing harm of the other. The question whether the utility of these side effects outweighs the utility of affecting people in the fold interval is a question of the overall utility of the gerrymanderer.

\subsection{Discussion}

In this section, I have shown that gerrymandering an IF predictor is possible, under certain assumptions about the structure of the predictor, the distribution of different groups, and also the utility of the gerrymanderer. Note that while the examples given above depend on the specific context (Euclidean metric on the real numbers), the examples may generalize to other metric spaces, because they use special cases of maps on generic metric spaces (isometries and contractions).

Gerrymandering IF is different than gerrymandering group fairness measures in that IF gerrymandering is more fine-grained. Specific assumptions have to be met for IF gerrymandering because IF depends on the metrics. Still, some instances of gerrymandering IF would be prevented by imposing group fairness measures. To give an example, gerrymandering a score through translation may violate statistical parity because translation may change the overall distribution of predictions.

The above examples of IF gerrymandering also bring a general feature of the very idea of individual fairness to the fore: IF is a \emph{relative} notion of fairness in that it is not defined in terms of the overall distribution of predictions, but only in terms of the relations between individuals. One drawback of such a relative notion is that the extent to which gerrymandering is ``worth it'' depends on the utility of the gerrymanderer: On the one end of the extreme, if a gerrymanderer has a preference of not affecting a particular group $B$, and if the different groups are already evenly distributed, then IF provides a good protection against gerrymandering for all groups because gerrymandering would affect $B$ as well. On the other end of the extreme, if a gerrymanderer wants to hurt group $A$ and does not care much about adversely affecting other groups by targeting one, or if groups are very unevenly distributed, then IF may not provide much protection against gerrymandering.

\section{Gerrymandering Metrics and Features}
\label{sec_metric_features}

\subsection{Idea}

In this section, I explore how the choice of the metrics $d$ and $D$ and feature space $X$ of individuals affects IF in particular contextss. This will reveal the extent to which metrics and feature space can be chosen by an attacker to yield unfavorable outcomes for a particular group, or for certain individuals.

An important difference between choosing the metrics and feature space on the one hand and gerrymandering the predictor as explored in the previous section on the other, is that choosing the metric $d$ between individuals and thereby obtaining different requirements on predictors is an intended feature of the notion of individual fairness: the metric $d$ is supposed to capture ``ground truth'' with respect to what a society considers to be fair, according to \citet[p. 1]{dwork2012}. Dwork et al. also write that the metric is supposed to be ``open to discussion and continual refinement'' (Ibid.). Only by choosing an appropriate metric $d$ does IF become a substantive notion of fairness. We will see that a lot hinges on the choice of metrics $d$ and $D$ and feature space $X$. The upshot of the following investigation is that if we do not place substantive restrictions on these choices, the resulting notion of IF is very permissive, or almost empty.

\subsection{Trivial and Discrete Metrics}

A first choice of (pseudo-)metric $d$ is the trivial metric. The trivial metric is identically zero, i.e., $d(p,q) = 0$ for all $p, q \in X$. With this metric, we have to choose a constant predictor $M(x) = c$ for all $x \in X$: If we choose different predictions $M(p) \neq M(q)$ for two individuals $p \neq q$, IF is violated. This means that by choosing a trivial metric, we can enforce the choice of a constant predictor. This choice may be unfair in some contexts; see the discussion in section \ref{sec:non-expansive} above.

A second choice is the discrete metric, i.e., $d(p, q) = 0$ if $p = q$, $d(p, q) =1$ if $p \neq q$. The substance of the notion of fairness resulting from this choice depends on the properties of the feature space $X$. One important property is the expressivity, or size, of $X$. We can choose a very large space $X$, which allows us to distinguish any two individuals. One way of implementing this is by including a unique identifier for all individuals, such as a representation of their complete DNA, pictures of their faces, fingerprints, and so on. If a feature space $X$ sufficiently large to represent such unique identifiers is combined with the discrete metric $d$, it is easy to see that the following proposition holds:

\begin{proposition}
Let $X$ be a feature space which allows for unique identifiers, $d$ the discrete metric, and the metric $D$ be normalized, i.e., $D(M(p), M(q) \leq 1$ for all $p, q \in X$. Then \emph{any} function $M:X \rightarrow \hat{Y}$ satisfies individual fairness.
\end{proposition}

To see this, note that in a feature space with unique identifiers, no two different individuals $p, q$ with $p \neq q$ are alike, which implies that $d(p, q) =1$, while $D(M(p), M(q) \leq 1$ for any $M$, such that any $M$ satisfies IF. Intuitively speaking, this means that with these choices of $X, d, D$, IF is an empty requirement: It is compatible with all possible predictors, which, of course, means that gerrymandering the predictor becomes very easy: A gerrymanderer can simply choose whatever predictor maximizes their utility.

If we choose $X$ to be very small, say, $X = \{0, 1\}$, we can choose to either set $0$ and $1$ to the same value in $\hat{Y}$, or we can choose to treat them differently. Intuitively, this means that we end up with two groups of people, those with feature $0$ and those with feature $1$, which can either be assigned the same or different predictions. If we choose $X$ to be a singleton set, i.e., if all individuals have the same features, we are forced to choose the constant predictor.

There are several lessons to be learned from the above cases. The choice of a maximally individualizing feature space $X$, combined with the discrete metric $d$ corresponds to an individualized notion of fairness, what \citet{binns2020} calls \emph{individualized justice}: All individuals have something that distinguishes them from all other people, and these distinguishing features can be used as a justification for treating them differently. This leads to a notion of fairness that does not allow for (statistical) generalization, but requires predictions and decisions on an individualized basis, as witnessed by the fact that there are no formal restrictions on the choice of predictor. This contradicts the claim by \citet[Sec. 4.2.]{binns2020} that IF is not compatible with individualized justice: If the feature space $X$ is sufficiently rich, IF draws distinctions between any two individuals, which turns IF into a kind of individualized justice.

The choice of a maximally discriminative feature space also shows that if IF does not come with additional requirements, it is maximally permissive by being compatible with any predictor, and thus constitues an empty requirement, a complaint usually directed against Aristotle's notion of \emph{consistency}, which is closely related to IF, cf. \citet[Sec. 3.1.]{binns2020}; \citet{fleish2021}. The choice of $X$ as maximally non-discriminative corresponds to a minimally individualized notion of fairness: If we all have the same features, ``We're all alike'' \citep{fried2016} and thus should be treated similarly (or equally).

What the above cases show is that IF in and of itself does not guarantee that fairness holds in any substantive sense. If one wishes to turn IF into a substantive notion of fairness, one has to put suitable restrictions on the choice of not only $M$,  $d$ and $D$, but also on the feature space $X$. Note that the importance of the choice of the feature space has been emphasized before in the literature; see, e.g., \citet{fried2016}.

\subsection{Absolute Individual Fairness}

In this section, I propose a special kind of IF, called \emph{absolute individual fairness}, a generalization of the results from the last section. Intuitively speaking, absolute individual fairness requires that if two individuals $p$ and $q$ have the same distributions of the ground truth $Y$ (e.g., probability of defaulting on a credit), then they have the same distribution of predictions $\hat{Y}$ (e.g., predicted probability of defaulting on a credit).

The predictors that satisfy absolute IF can thus be considered to approximate a probabilistic function from $X$ to $Y$, similar to the standard setting of supervised learning. In the present section , I assume that the predictor $M$ predicts \emph{distributions} over $\hat{Y}$; it is a probabilistic function $M:X \rightarrow \Delta(\hat{Y})$. Define $f_Y(x) := P(Y|X =x)$, and $f_{\hat{Y}}(x) := P(\hat{Y}|X = x)$. 

\begin{definition}
Let $d$ be the discrete metric on $f_Y(.)$ and $D$ the discrete metric on $f_{\hat{Y}}(.)$ A probabilistic predictor $M:X \rightarrow \Delta(\hat{Y})$ satisfies absolute individual fairness with respect $Y$ if for all $p, q$ in $X$, it holds: if $f_Y(p) = f_Y(q)$, then $f_{\hat{Y}}(p) = f_{\hat{Y}}(q)$. 
\end{definition}

It is not hard to see that absolute individual fairness is a kind of individual fairness, on the assumption that on $X$ we use the discrete metric induced by $f_Y(.)$: Individuals are identified, i.e., have distance $0$ iff. they have the same distribution of ``ground truths''. With these choices, IF would be violated if, for some $p, q \in X$, $d(p,q) = 0$ and $D(p,q) = 1$. This would imply $f_Y(p) = f_Y(q)$ and $f_{\hat{Y}}(p) \neq f_{\hat{Y}}(q)$, which is not possible due to absolute individual fairness.

To better understand what absolute individual fairness means, it is useful to have a different characterization. This characterization is based on the fact that the two discrete metrics $d$ and $D$ partition the space $X$: they group all and only those individuals in $X$ together that have the same distribution with respect to the ground truth and the prediction, respectively. On this basis, one can prove that a predictor satisfies absolute individual fairness if and only if the partition of $X$ induced by the predictor is more coarse-grained than the partition of $X$ induced by the ground truth. To describe the partitions, one needs the notion of a minimal sufficient statistic, which characterizes the coarsest partition of a space that groups together those elements that have the same distribution with respect to a different variable. One can then prove the following proposition; see appendix \ref{sec:aif_proof} for formal definitions and proofs:

\begin{proposition}
A probabilistic predictor $M:X \rightarrow \Delta(\hat{Y})$ satisfies absolute individual fairness with respect to ground truth $Y$ iff. the minimal sufficient statistic $U(X)$ of $X$ for $f_{\hat{Y}}$ is a function of the minimal sufficient statistic $T(X)$ of $X$ for $f_Y$. 
\label{prop:aif}
\end{proposition}

This means that whether or not a predictor satisfies absolute individual fairness depends on how fine-grained the partition of $X$ with respect to $f_Y$ is. If this partition is very fine-grained, then only few individuals have the same distributions of ground truth. And if this is the case, then more predictors will satisfy absolute individual fairness, because the predictor only needs to preserve a very fine partition. If this partition is coarse-grained, then many individuals have the same ground truth. In this case, the requirements on the predictor are more stringent.

Now, the key question is: What kind of partition of $X$ with respect to $f_Y$ can we expect? Will this partition usually be fine-grained or coarse-grained? There is a precise answer to this question in the statistical setting. The so-called Pitman-Koopman-Darmois theorem tells us that if the random variable $X$ does not follow a particularly nice distribution (from the exponential family; see \citet{casel2002} for details), then the minimal sufficient statistic $T(X)$ of $X$ with respect to $f_Y$ will not be much more coarse-grained than $X$ itself. This, in turn, means that if $X$ captures fine-grained information about individuals, as will be the case if $X$ is high-dimensional and ``individualized'', then the minimal sufficient statistic of $X$ with respect to $f_Y$ will be very fine-grained, and thus not provide much of a restriction on fair predictors. This generalized the lesson we already learned in the previous section: If many feature of individuals are taken into account, individual fairness is a very weak requirement.

\subsection{Discussion}

In this section, we have seen that IF strongly depends on the kind of metrics and feature space $X$ that is employed. In particular, if $X$ is very fine-grained, this can be used as a justification to essentially treat all people individually, as one pleases, while still formally satisfying IF. If $X$ is very coarse grained, then this forces us to use a predictor that takes the same value for many people. This dependence on properties of $X$ carries over from a setting with point predictions to a probabilistic setting with distributions over predictions. This shows that, as a formal requirement, IF is very weak, not providing any restrictions on possible predictors in some cases.

It could be thought that a lot hinges on the use of the discrete metric, and it could be wondered whether the insights in this section are due to particular features of the discrete metric. The main feature of the metric that yields the phenomena we saw here is the weight it puts on features that distinguish different individuals. For the discrete metric, these weights yield categorical differences. However, it is also possible to construct metrics that draw more quantitative distinctions, emphasizing different features of individuals differently. The property of a metric that matters is how much weight it puts on individualizing features of individuals, which allows them to be treated more or less differently.

\section{Leibniz Fairness}
\label{sec:leibnitz}

In this section, I consider two questions. First, what are the main problems and advantages of IF, generally speaking? Second, is it possible to improve on IF, by retaining the advantages of IF, while avoiding the problems?

First, let me discuss the problems of IF. IF allows for gerrymandering by manipulating the predictor, and by choice of metric $d$ and feature space $X$. The reason for these problems is that IF is formulated as a \emph{metric} requirement, which is too weak, for example in a setting where predictions have an order structure, which is not preserved by a metric requirement, as shown in section \ref{sec:ger_pred}. The fundamental problem with a metric formulation of fairness is that it is just not clear why what matters for fairness is \emph{metric} structure. We have seen in section \ref{sec_metric_features} that, as a formal requirement, IF is not much of a restriction on the choice of predictors, at least for some choices of metric and feature space. It could be thought that this is beneficial in that IF presupposes ``fairness through awareness'', i.e., it forces us to make our choices of fairness explicit. But this does not resolve the issue: It is equally unclear why it is a good idea to ``express our awareness'' of fairness through the metric structure of a prediction problem. Note that IF has other problems that have been pointed out in the literature, such as the problem of incommensurability pointed out by \citet{fleish2021}, viz. that IF requires that there need to be distances between all individuals. This can also be seen as a problem of the metric formulation of fairness.

Turning to the advantages of IF. First, IF is able to identify some cases of fairness gerrymandering that occur at the sub-group level. IF thus complements group fairness measures. Second, the general idea of ``fairness through awareness'' also has some merit. The idea that we should make our choices of fairness explicit by codifying them in some form and posing some restrictions on possible predictors seems sound.

Turning to the second question, is it possible to retain the advantages of IF as much as possible, while avoiding the problems? In order to solve the problem with gerrymandering, one possibility is to formulate requirements on the individual level that are stronger than metric conditions. For example, if a prediction problem has order structure, one requirement could be that the order structure should be preserved on top of the metric structure, thus preventing some kinds of fairness gerrymandering. On this approach, IF would be fixed in a piecemeal manner, depending on the structure of the prediction problem at hand. This, however, does not provide us with a general recipe for approaching fairness, because the approach is local and piecemeal. Not all prediction problems have the same structure, and it would be desirable to have a more general starting point.

To have such a general starting point, let us consider the question of how we would solve prediction problems under ideal circumstances, without thinking about practical matters. How would we construct fair predictors if we had unlimited resources to determine fair predictors? We may want to take both features of individuals as well as group membership into account; the ultimate choice of predictor may depend on both. We may want the outcome to depend on merit (or ground truth), but not necessarily. The form of prediction may be real valued, or have less structure, e.g. a partial order, so as to make room for the possibility of incommensurability, or abstention of prediction. Finally, there are different sources or ways of grounding fairness. There are moral intuitions, which are defeasible, and there are different kinds of moral principles, which are contested. The latter may include different restrictions on predictors due to group membership or individual merit.

To carry out this ideal approach, we have to first settle for the appropriate variables, or spaces, of the prediction problem. We need a variable $X$ characterizing individuals. Other variables, such as $A$ for group membership, may be functions of $X$, or they may be independent random variables. Further variables are ground truth or ``merit'' $Y$, and predictions $\hat{Y}$. Importantly, the specification of the properties of these variables is part and parcel of the resulting notion of fairness. For example, if we want ``individualized justice'', i.e., predictions that depend on the features of individuals , then $X$ needs to be sufficiently fine-grained to represent unique identifiers for individuals. The same goes for $Y$ and $\hat{Y}$, which may need to be appropriately structured to make room for abstention, or incommensurable predictions, such that, e.g., $\hat{Y}$ is a partial orders; see \cite{fleish2021} for more on why this may be desirable.

Second, we have to come up with a fair predictor. Given a finite set of individuals, represented by $X_0 \subset X$, a fair predictor $M$ would be constructed by determining, for each $x \in X_0$, a probability distribution over $\hat{Y}$. The distribution $f_{\hat{Y}}(x)$ for $x \in X_0$ can depend on the group membership $A$ of each $x$, the merit (or ground truth) $Y$, and on further, individual features. The specification of a distribution $f_{\hat{Y}}$ for each $x \in X_0$ could be called a \emph{Leibniz fair} predictor, because it is the most stringent way of saying what a fair predictor is -- it is the fairest predictor of all possible worlds, cf. \cite{look2020}. All possible tradeoffs between properties of individuals and their group membership have to be taken into account to determine all distributions, such that violations of measures of group fairness are by choice. Specifying the full distribution of predictions for all individuals in $X_0$ prevents gerrymandering because the (distribution of) predictions are rigid. There is no formal measure of fairness that leaves any leeway for manipulation. Of course, the determination of distributions over $\hat{Y}$ may be based on both moral intuitions, ethical principles, empirical knowledge, and mixtures thereof. To make an analogy with cryptography, Leibniz fairness is to fairness what a one-time-pad is to secure encryption.

Note that Leibniz fairness is not a formal criterion to verify whether or not a given predictor is fair. Also, it does not tell us anything about novel predictions. This is because if we want to make predictions for an individual $x^*$ not represented in $X_0$, then, in the best of all possible worlds, we have to start anew with a set $X_1 = X_0 \cup \{x^* \}$ and find a Leibniz fair predictor for this new set. This is necessary because the addition of the individual $x^*$ may change how we think other people should be treated, and thus affect the distributions of others as well. 

How, then, is Leibniz fairness useful? It is useful because it provides an ``ideal'' starting point to propose more feasible measures of fairness, which take both individuals and aggregates of individuals into account. Instead of investigating how to approximate, say, IF through elicitation (see references in Sec. \ref{sec:background}), it may be more adequate to investigate how Leibniz fairness can be approximated, say, in particular settings, which also makes it possible to explicitly address the question whether deviations from Leibniz fairness are justifiable under particular circumstances.

\section{Conclusion}

The present paper argued that it is possible to gerrymander predictors that satisfy individual fairness in particular contexts and under certain circumstances. In particular, gerrymandering is possible if different, socially salient groups are unevenly distributed over the prediction space. This makes it possible to implement local fairness attacks against these groups. The paper also argued that in different contexts and for certain choices of metrics and feature spaces, individual fairness is a very weak fairness requirement. Finally, the paper proposed a stronger notion of fairness, Leibniz fairness, a non-formal notion of fairness that provides the strongest possible protection against gerrymandering.

\bibliographystyle{natbibeng}
\bibliography{bibliography_gif}

\begin{thebibliography}{23}
\providecommand{\natexlab}[1]{#1}

\bibitem[{Altman(2020)}]{altma2020}
Altman, A. 2020.
\newblock {Discrimination}.
\newblock In E.~N. Zalta, ed., \emph{The {Stanford} Encyclopedia of
  Philosophy}. Metaphysics Research Lab, Stanford University, {W}inter 2020 ed.

\bibitem[{Angwin et~al.(2016)Angwin, Larson, Mattu, and Kirchner}]{angwi2016}
Angwin, J., J.~Larson, S.~Mattu, and L.~Kirchner. 2016.
\newblock Machine bias: There's software used across the country to predict
  future criminals. and it's biased against blacks.
\newblock ProPublica.

\bibitem[{Barocas et~al.(2019)Barocas, Hardt, and Narayanan}]{baroc2019}
Barocas, S., M.~Hardt, and A.~Narayanan. 2019.
\newblock \emph{Fairness and Machine Learning}.
\newblock fairmlbook.org.

\bibitem[{Binns(2018)}]{binns2018}
Binns, R. 2018.
\newblock Fairness in machine learning: Lessons from political philosophy.
\newblock In \emph{Conference on Fairness, Accountability and Transparency}.
  PMLR, pp. 149--159.

\bibitem[{Binns(2020)}]{binns2020}
---{}---{}---. 2020.
\newblock On the apparent conflict between individual and group fairness.
\newblock In \emph{Proceedings of the 2020 conference on fairness,
  accountability, and transparency}. pp. 514--524.

\bibitem[{Casella and Berger(2002)}]{casel2002}
Casella, G. and R.~L. Berger. 2002.
\newblock \emph{Statistical Inference}.
\newblock Duxbury, second ed.

\bibitem[{Dwork et~al.(2012)Dwork, Hardt, Pitassi, Reingold, and
  Zemel.}]{dwork2012}
Dwork, C., M.~Hardt, T.~Pitassi, O.~Reingold, and R.~S. Zemel. 2012.
\newblock Fairness through Awareness.
\newblock Proc. ACM ITCS, pp. 214---226.

\bibitem[{Fleisher(2021)}]{fleish2021}
Fleisher, W. 2021.
\newblock What's Fair about Individual Fairness?
\newblock In \emph{Proceedings of the 2021 AAAI/ACM Conference on AI, Ethics,
  and Society}. New York, NY, USA: Association for Computing Machinery, pp.
  480--490.

\bibitem[{Friedler et~al.(2016)Friedler, Scheidegger, and
  Venkatasubramanian}]{fried2016}
Friedler, S.~A., C.~Scheidegger, and S.~Venkatasubramanian. 2016.
\newblock On the (im) possibility of fairness.
\newblock ArXiv:1609.07236.

\bibitem[{Ilvento(2019)}]{ilven2019}
Ilvento, C. 2019.
\newblock Metric learning for individual fairness.
\newblock \emph{arXiv preprint arXiv:1906.00250} .

\bibitem[{Jung et~al.(2019)Jung, Kearns, Neel, Roth, Stapleton, and
  Wu}]{jung2019}
Jung, C., M.~Kearns, S.~Neel, A.~Roth, L.~Stapleton, and Z.~S. Wu. 2019.
\newblock An algorithmic framework for fairness elicitation.
\newblock \emph{arXiv preprint arXiv:1905.10660} .

\bibitem[{Kearns et~al.(2018)Kearns, Neel, Roth, and Wu}]{kearn2018}
Kearns, M., S.~Neel, A.~Roth, and Z.~S. Wu. 2018.
\newblock Preventing Fairness Gerrymandering: Auditing and Learning for
  Subgroup Fairness.
\newblock \emph{PMLR} 80: 2564--2572.

\bibitem[{Kearns and Roth(2019)}]{kearn2019}
Kearns, M. and A.~Roth. 2019.
\newblock \emph{The ethical algorithm: The science of socially aware algorithm
  design}.
\newblock Oxford University Press.

\bibitem[{Look(2020)}]{look2020}
Look, B.~C. 2020.
\newblock {Gottfried Wilhelm Leibniz}.
\newblock In E.~N. Zalta, ed., \emph{The {Stanford} Encyclopedia of
  Philosophy}. Metaphysics Research Lab, Stanford University, {S}pring 2020 ed.

\bibitem[{Mitchell et~al.(2021)Mitchell, Potash, Barocas, D'Amour, and
  Lum}]{mitch2021}
Mitchell, S., E.~Potash, S.~Barocas, A.~D'Amour, and K.~Lum. 2021.
\newblock Algorithmic fairness: Choices, assumptions, and definitions.
\newblock \emph{Annual Review of Statistics and Its Application} 8: 141--163.

\bibitem[{Mukherjee et~al.(2020)Mukherjee, Yurochkin, Banerjee, and
  Sun}]{mukhe2020}
Mukherjee, D., M.~Yurochkin, M.~Banerjee, and Y.~Sun. 2020.
\newblock Two simple ways to learn individual fairness metrics from data.
\newblock In \emph{International Conference on Machine Learning}. PMLR, pp.
  7097--7107.

\bibitem[{Petersen et~al.(2021)Petersen, Mukherjee, Sun, and
  Yurochkin}]{peter2021}
Petersen, F., D.~Mukherjee, Y.~Sun, and M.~Yurochkin. 2021.
\newblock Post-processing for Individual Fairness.
\newblock \emph{Advances in Neural Information Processing Systems} 34.

\bibitem[{Petrunin(2021)}]{petru2021}
Petrunin, A. 2021.
\newblock Euclidean Plane and its Relatives.
\newblock Https://math.libretexts.org/@go/page/23576 (accessed 2022-01-26).

\bibitem[{R{\"a}z(2021)}]{raez2021}
R{\"a}z, T. 2021.
\newblock Group Fairness: Independence Revisited.
\newblock In \emph{Proceedings of the 2021 ACM Conference on Fairness,
  Accountability, and Transparency}, FAccT '21. New York, NY, USA: Association
  for Computing Machinery, pp. 129--137.

\bibitem[{Sharifi-Malvajerdi et~al.(2019)Sharifi-Malvajerdi, Kearns, and
  Roth}]{shari2019}
Sharifi-Malvajerdi, S., M.~Kearns, and A.~Roth. 2019.
\newblock Average individual fairness: Algorithms, generalization and
  experiments.
\newblock \emph{Advances in Neural Information Processing Systems} 32.

\bibitem[{Vargo et~al.(2021)Vargo, Zhang, Yurochkin, and Sun}]{vargo2021}
Vargo, A., F.~Zhang, M.~Yurochkin, and Y.~Sun. 2021.
\newblock Individually fair gradient boosting.
\newblock \emph{arXiv preprint arXiv:2103.16785} .

\bibitem[{Yurochkin et~al.(2019)Yurochkin, Bower, and Sun}]{yuroc2019}
Yurochkin, M., A.~Bower, and Y.~Sun. 2019.
\newblock Training individually fair ML models with sensitive subspace
  robustness.
\newblock \emph{arXiv preprint arXiv:1907.00020} .

\bibitem[{Yurochkin and Sun(2020)}]{yuroc2020}
Yurochkin, M. and Y.~Sun. 2020.
\newblock Sensei: Sensitive set invariance for enforcing individual fairness.
\newblock \emph{arXiv preprint arXiv:2006.14168} .

\end{thebibliography}

\appendix

\section{Proof of Proposition \ref{prop:folding}}
\label{sec:folding_proof}

The folding $\phi:\mathbb{R} \rightarrow \mathbb{R}$ for an interval $[a, b] \subset \mathbb{R}$, $a < b$ is defined as

\begin{equation*}
\phi(x) = \begin{cases}
x - 2(b-a) & \text{if } x > b, \\
-x + 2a & \text{if } x \in [a, b] \\
x & \text{if } x < a.
\end{cases}
\end{equation*}

We have to prove that the folding map is non-expansive, i.e., that for all $p, q \in \mathbb{R}$, it holds $|\phi(p) - \phi(q)| \leq |p - q|$. W.l.o.g. we assume $p > q$. There are six cases to consider. All inequalities are estimated with the triangle inequality.

Cases 1-3, $p, q > b$; $p, q \in [a, b]$; $p, q < a$: in these cases, $\phi$ is an isometry, and thus non-expansive.

Case 4, $p > b, q \in [a, b]$: $|\phi(p) - \phi(q)| = |p - 2(b-a) - (-q + 2a)| = |p + q - 2b|$. If $q =b$, we have  $|p + q - 2b| = |p - q|$ and we are done. If $q < b$, we define $\epsilon, \delta > 0$ with $p = b + \epsilon$, and $q = b - \delta$. We continue $|p + q - 2b| = |b - \delta + b + \epsilon - 2b| = |\epsilon - \delta| \leq |\epsilon + \delta| = |\epsilon + b - b + \delta | = |p - q|$.

Case 5, $p > b, q < a$: $|\phi(p) - \phi(q)| = | p - 2 (b -a) -q | = |p - q - 2(b-a)| $. Define $\epsilon, \delta > 0 $ with $p = b + \epsilon$ and $q = a - \delta$. We continue $|p - q - 2(b-a)| = |b + \epsilon - a + \delta - 2 (b-a)| = |\epsilon + \delta - (b-a)| \leq  |\epsilon + \delta + (b - a)| = |p - b + a - q + (b-a)| = |p-q|$.

Case 6: $p \in [a, b], q < a$: This case hold in virtue of symmetry with case 4. For completeness's sake, here is the proof: $|\phi(p) - \phi(q)| = |- p + 2a - q|$. If $p = a$, we have $|-p + 2a - q| = |p - q|$ and we are done. If $p > a$, we define $\epsilon, \delta > 0$ with $p = a + \epsilon$, $q = a - \delta$. We continue $|- p + 2a - q| = |- a - \epsilon + 2a - a + \delta| = |\delta - \epsilon| \leq |\delta + \epsilon| = |p - a + a - q| = |p-q|$.

\section{Proof of Proposition \ref{prop:aif}}
\label{sec:aif_proof}

This appendix provides a proof of proposition \ref{prop:aif}. First, some standard definitions.

\begin{definition}
Let $X, Y$ be random variables and $S$ a statistic (function) of $X$. $S$ is a \emph{sufficient statistic} of $X$ for $Y$ if $P(Y \mid X, S(X)) = P(Y \mid S(X))$.
\end{definition}

This means that, for the purpose of predicting $Y$, we do not need the full $X$, we only need $S(X)$.

\begin{definition}
Let $X, Y$ be random variables and $T$ a sufficient statistic of $X$ for $Y$. $T$ is a \emph{minimal sufficient statistic} of $X$ for $Y$ if it is a function of every other sufficient statistic, i.e., if, for every other sufficient statistic $S(X)$ of $X$ for $Y$, there exists a function $f$ such that $T(X) = f(S(X))$.
\end{definition}

This means that, of all the statistics that are sufficient for $Y$, $T(X)$ contains the least amount of information. We now turn to the proof of proposition \ref{prop:aif}:

``$\Rightarrow$'': Assume that a predictor $M:X \rightarrow \Delta(\hat{Y})$ satisfies absolute individual fairness with respect to $Y$. We have to show that the minimal sufficient statistic $U(X)$ of $X$ with respect to $f_{\hat{Y}}$ is a function of the minimal sufficient statistic $T(X)$ of $X$ with respect to $f_Y$. Assume that $T(p) = T(q)$ for some $p, q$. This implies that $f_Y(p) = f_Y(q)$, because $T(X)$ is the minimal sufficient statistic of $X$ with respect to $f_Y$. The definition of absolute individual fairness implies that $f_{\hat{Y}}(p) = f_{\hat{Y}}(q)$. Now if $U(X)$ is the minimal sufficient statistic of $X$ with respect to $f_{\hat{Y}}$, this also implies $U(p) = U(q)$. Otherwise, $U(X)$ would send elements of $X$ with the same distribution to different values, a contradiction with it being a m.s.s. . Thus, $U(X)$ is a function of $T(X)$.

``$\Leftarrow$'': Assume that $U(X)$, the m.s.s. of $X$ with respect to $f_{\hat{Y}}$, is a function of $T(X)$, the m.s.s. of $X$ with respect to $f_Y$. Assume that $p, q$ are elements of $X$ that satisfy $f_Y(p) = f_Y(q)$. Now, because $T(X)$ is the m.s.s. of $X$ with respect to $f_Y$, this implies that $T(p) = T(q)$. By assumption, there is a function $g$ such that $U(X) = g(T(X))$. We thus get $g(T(p)) = g(T(q))$ and thus $U(p) = U(q)$. This, in turn, implies that $f_{\hat{Y}}(p) = f_{\hat{Y}}(q)$, which means that the predictor $M:X \rightarrow \Delta(\hat{Y})$ satisfies absolute individual fairness with respect to $Y$.

\end{document}